\documentclass[conference]{IEEEtran}
\usepackage{cite}
\usepackage{amsmath,amssymb,amsfonts}
\usepackage{algorithmic}
\usepackage{graphicx}
\usepackage{amsthm}
\newtheorem*{remark}{Remark}
\usepackage{multirow}
\ifCLASSOPTIONcompsoc
\usepackage[caption=false,font=normalsize,labelfont=sf,textfont=sf]{subfig}
\else
\usepackage[caption=false,font=footnotesize]{subfig}
\fi
\usepackage{stfloats}
\usepackage{textcomp}
\usepackage{xcolor}
\begin{document}

\title{Design of $\mathcal{H}_{\infty}$-based Robust Controller for Single-phase Grid-feeding Voltage Source Inverters}

\author{\IEEEauthorblockN{Soham Chakraborty,
Sourav Patel, and Murti V. Salapaka}
\IEEEauthorblockA{Department of Electrical and Computer Engineering\\
University of Minnesota Twin Cities
\\ \{chakr138, patel292, murtis\}@umn.edu}}

\maketitle
\begin{abstract}
This article proposes the design of $\mathcal{H}_{\infty}$-based robust current controller for single-phase grid-feeding voltage source inverter with an $LCL$ filter. The main objective of the proposed controller is to have good reference tracking, disturbance rejection and sufficient $LCL$ resonance damping under a large range of variations of grid impedance. Based on the aforementioned performance requirements, frequency dependent weighting functions are designed. Subsequently, the  sub-optimal control problem is formulated and solved to determine the stabilizing controller. Computational footprint of the controller is addressed for ease-of-implementability on low-cost controller boards. Finally controller hardware-in-the-loop simulations on OPAL-RT are performed in the validation stage to obtain performance guarantees of the controller. The proposed controller exhibits fast response during transients and superior reference tracking, disturbance rejection at steady-state when compared with proportional- and resonant-based current controllers.
\end{abstract}

\begin{IEEEkeywords}
$\mathcal{H}_{\infty}$-based loop shaping, current controller, parametric uncertainty, robust control, voltage source inverter.
\end{IEEEkeywords}

\section{Introduction}
With the proliferation of distributed energy resources (DERs), grid-feeding voltage source inverters (\textit{gf}VSIs) have become an essential component of the distribution network. These VSIs are usually connected to a network with regulated voltage and frequency maintained either by the grid or by the local grid-forming DERs \cite{DER_control_Madureira}. \textit{gf}VSIs are often terminated by $LCL$ filters and operated with a current control scheme in order to inject a regulated current into the network with a desired power factor and limited harmonic content \cite{filter}.
\par Various types of control schemes and their advancements for \textit{gf}VSIs have been proposed in literature. Classical controllers such as proportional-integral (PI) controller-based control in $d$-$q$ domain, proportional-resonant (PR) controller-based control in $\alpha$-$\beta$ domain for \textit{gf}VSIs are most popular due to the simplicity in design and ease of implementation \cite{pipr2,pipr5,yazdani}. However, lack of robustness in performance of these controllers due to varying system conditions are one of the major drawbacks \cite{piprmod2}. Hysteresis current controllers are equally popular due to the advantage of simplicity and robustness \cite{hyst1}. However, the major limitation of hysteresis control is the dependency of the VSI switching frequency on the load parameters resulting in degraded performance with current harmonics ripple \cite{modelpre1}. Model-predictive current controllers are proposed in \cite{modelpre1} in order to circumvent these limitations. However, powerful controller platform is a prerequisite to employ this control scheme due to their added complexity. Other advanced controller designs e.g. linear quadratic regulator-based full-state feedback controller \cite{statefeed3}, sliding-mode controller \cite{sliding1}, repetitive controller \cite{repet1}, two degree-of-freedom quasi-PI controller \cite{twodegree1}, are also proposed at the cost of controller complexity and added computational burden. Moreover, damping of $LCL$ filter resonance is not considered in these works that can lead to performance degradation and stability issue. Damping of $LCL$ resonance is realized either actively by including additional feed-forward loops of the voltage and current measurements from the filter as proposed in \cite{lcldamp2} or passively by proposing new types of output filters \cite{newfilter}. Moreover, reference \cite{robustcon1} proposes a state feedback as active damping for the $LCL$ filter resonance. Reference \cite{hornic1} provides a repetitive control for current controller of grid-connected VSIs with no $LCL$ resonance damping scheme.
\par Most of the aforementioned controllers lack robustness in performance with a wide-range of variations in grid impedance parameters that exists due to ever changing adjustments of the network configurations through opening and closing of circuit breakers. The inevitable variation in equivalent grid impedance (experienced by the \textit{gf}VSIs) has a large impact on the resultant $LCL$ resonant frequency and as a result on the stability \cite{lcldamp6}. Reference \cite{gridimp4} proposes a dual-loop current control scheme with grid-side inverter current and capacitor current measurement feedback system to damp the resonance effect. Enhanced transient response and robustness against the grid impedance variations are achieved at the cost of increased number of sensors for the controller, hence increase in the cost. 
\par Robust active damping of the varying resonance and stability of the control system with grid impedance uncertainties are the greater concerns for \textit{gf}VSIs today. Robust design of controllers with performance criteria is gaining attention recently. Reference \cite{gridimp2} proposed a robust outer-filter inductor current controller along with classical PI-based inner-filter inductor current controller. This architecture introduces robustness of the controller in presence of uncertain grid impedance at the cost of increased number of current sensors. Moreover, active damping of filter resonance is not considered in the design. 
\par This article presents a design of $\mathcal{H}_{\infty}$-based optimal current controller for single phase \textit{gf}VSIs with good reference tracking, disturbance rejection, sufficient $LCL$ resonance damping and is robust under a pre-specified range of variations of grid impedance, that is not well-studied in the current literature. Controller hardware-in-the-loop (CHIL) based real-time simulation on OPAL-RT are performed for testing the viability and efficacy of the proposed controller. The proposed architecture is compared against conventional PR-based controllers and are shown to have superior performance in attaining the above-mentioned objectives in presence of uncertain grid impedance parameters. The contribution of this article is toward achieving robust performance of the current controller against the variation of grid impedance. It also facilitates the control with low complexity and no additional requirements of sensors.
\begin{figure}[t]
	\centering
    \includegraphics[scale=0.22,trim={2.2cm 2.2cm 10.5cm 1.5cm},clip]{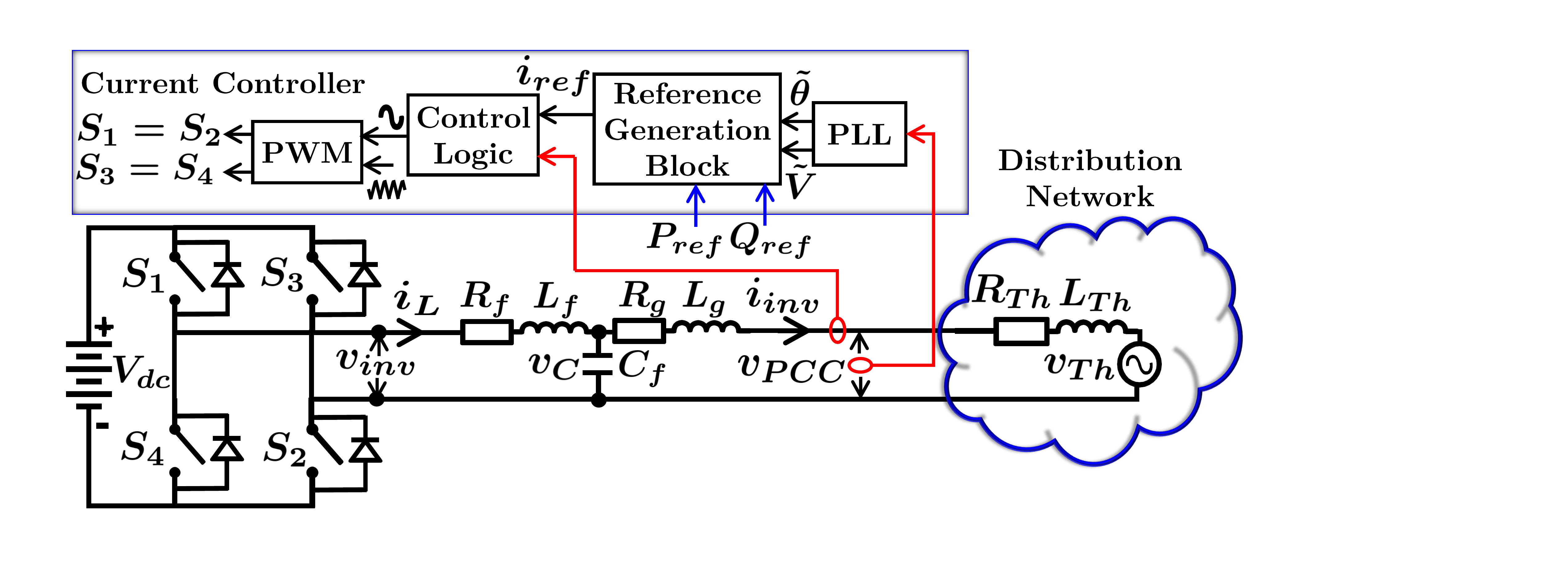}%
	\caption{A single-phase grid-feeding VSI connected to distribution network.}
	\label{fig:circuit}
\end{figure}
\begin{figure}[t]
	\centering
    \includegraphics[scale=0.26,trim={0cm 8.25cm 0cm 0cm},clip]{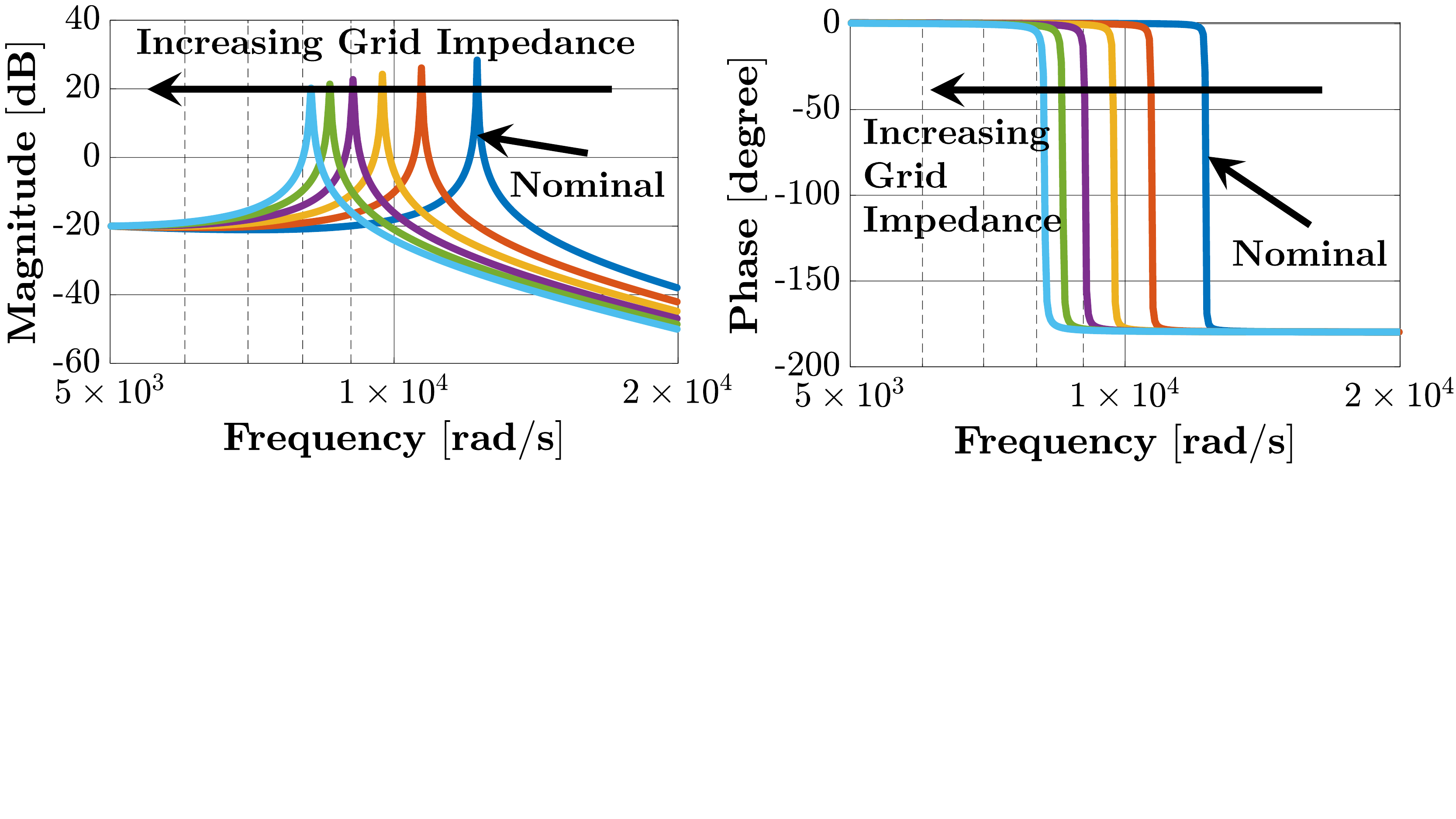}%
	\caption{Bode plot of $\mathcal{G}_{inv}$ with varying grid impedance.}
	\label{fig:resonant}
\end{figure}
\section{Plant Model Description of Grid-Feeding VSI}
The power circuit of a single-phase \textit{gf}VSI connected to a distribution network at the point of common coupling (PCC) is shown in Fig.~\ref{fig:circuit}. The VSI is composed of a dc bus, $V_{dc}$, four switching devices, $S_1, S_2, S_3, S_4$, and an $LCL$ filter with $L_f$, $L_g$ and $C_f$ as filter inductors and capacitor respectively with $R_f$ and $R_g$ as parasitic resistances of filter inductors. The distribution network is represented by the Thevenin equivalent voltage source, $v_{Th}$, in series with the Thevenin equivalent impedance $Z_{Th}:=R_{Th}+j\omega_oL_{Th}$ where $\omega_o$ (in rad/sec) is the nominal frequency of the distribution network. The VSI operates in $P$-$Q$ control mode by employing a current control strategy to regulate its output real and reactive power while being supported by a stable voltage and frequency source such as the grid, hence called grid-feeding VSI. The controller uses sinusoidal PWM switching technique to generate the switching signals. The dynamics of the VSI are described as:
\begin{align}
L_f \frac{d\langle i_{L}\rangle}{dt} +R_f\langle i_{L}\rangle &= \langle v_{inv}\rangle - \langle v_C\rangle, \label{eq1}\\
L_T \frac{d\langle i_{inv}\rangle}{dt} +R_T\langle i_{inv}\rangle &= \langle v_{C}\rangle - \langle v_{Th}\rangle, \label{eq2}\\
C_f \frac{d\langle v_C\rangle}{dt} &=  \langle i_{L}\rangle - \langle i_{inv}\rangle, \label{eq3}
\end{align}
where $\langle .\rangle$ signifies the average values of the corresponding variable over one switching cycle ($T_s$) \cite{yazdani}. Laplace transformation and algebraic manipulation with \eqref{eq1}, \eqref{eq2} and \eqref{eq3} result:
\begin{align}\label{eq4}
I_{inv}(s) &= \mathcal{G}_{inv}(s)V_{inv}(s) - \mathcal{G}_{Th}(s)V_{Th}(s),
\end{align}
where, $\mathcal{G}_{inv}(s)$ and $\mathcal{G}_{Th}(s)$ are transfer functions parameterized by $L_f$, $R_f$, $C_f$, $L_T:=L_g+L_{Th}$ and $R_T:=R_g+R_{Th}$.
\section{Impacts of Network on Plant Model Dynamics}
As observed in \eqref{eq4}, distribution network has impacts on the open-loop plant dynamics of a \textit{gf}VSI. Firstly, both $\mathcal{G}_{inv}(s)$ and $\mathcal{G}_{Th}(s)$ consist of $L_T$ and $R_T$ (in turn consists of $L_{Th}$ and $R_{Th}$) as parameters that introduce uncertainties in the plant model. Secondly, $V_{Th}(s)$, imposed by the network, is acting as an exogenous disturbance signal to the plant. Where the latter one is a classical disturbance rejection problem, the severity of the former one questions the robust performance and needs to be discussed elaborately which is presented next.
\subsection{Effect of Thevenin Equivalent Grid Impedance Parameters}
The resonant frequency of the $LCL$ circuit in Fig.~\ref{fig:circuit}, neglecting resistive elements of the circuit, is given as:
\begin{align}
    f_{res} = \dfrac{1}{2\pi}\sqrt{\dfrac{L_f+L_T}{L_fL_TC_f}}~(\text{in}~\text{Hz}).
\end{align}
The presence of parasitic resistances can provide passive damping to the resonance phenomenon, however minimal. Therefore, it is recommended to keep the pass-band of the current controller smaller than the resonant frequency, pre-determined based on the designed $L_f$, $L_g$ and $C_f$, in order to avoid instability due to resonance phenomenon. However, it is observed that the resultant resonant frequency is sensitive to grid parameters, especially grid inductance ($L_{Th}$). In case of a stiff grid with small $L_{Th}$, provided a properly designed filter, it can be ensured, to some extent, that the resultant resonant frequency is larger than the bandwidth of the controller. However, this issue is more severe in case of a weak grid system associated with large $L_{Th}$. For a sufficiently weak grid, the resultant resonance frequency may decrease and the corresponding resonance peak may enter the pass-band of the current controller as illustrated in Fig.~\ref{fig:resonant}. This uncertainty in grid impedance introduces difficulties in controller design \cite{lcldamp6}. In severe most situation, e.g. weak grid conditions, due to low-power transformers and long power lines in rural areas, this in turn results in instability of the \textit{gf}VSIs as evidenced in \cite{gridimp4}.
\subsection{Modeling of Uncertain Grid Impedance}
A systematic design of grid impedance variation for leveraging the \textit{gf}VSI control is elaborated in this section. The open-loop plant model for the controller, as given in \eqref{eq1}, \eqref{eq2}, \eqref{eq3}, of the circuit configuration in Fig.~\ref{fig:circuit} is shown in Fig.~\ref{fig:control1} (inside the blue box). Clearly, variations in grid impedance (i.e. $R_{Th}$ and $L_{Th}$) results in real-parametric uncertainties in the parameters, $R_T$, $L_T$, of the transfer function (highlighted in red) in Fig.~\ref{fig:control1}. The short-circuit ratio (SCR) is often used to characterize the grid stiffness/weakness and it comes handy in determining the equivalent impedance of the grid at the PCC \cite{lcldamp6}. It is mathematically defined as $(V_{PCC}^{Nom})^2/\big[S_{B}\sqrt{(\omega_oL_g)^2+R_g^2}\big]$,
where, $V_{PCC}^{Nom}$ is the nominal RMS voltage at PCC, and $S_B$ is the rated apparent power of the \textit{gf}VSI. Usually the grid at PCC is considered as weak when the SCR is less than $3$. In this work, with a pre-specified SCR ($< 3$) and given $X/R$ ratio of grid, the nominal grid impedance parameter are determined, denoted as $L^{N}_{Th}$ and $R^{N}_{Th}$. By considering $\pm 100\%$ variations over nominal values, it is assumed that $L_{Th}\in [\underline{L}_{Th},\bar{L}_{Th}]$ and $R_{Th}\in [\underline{R}_{Th},\bar{R}_{Th}]$. It is to be noted that very stiff to extremely weak grid conditions are considered by designing the grid impedance parameters in this way. As a result,
\begin{align}
    L_T := L_T^{N} + w_L\delta_L,~ R_T := R_T^{N} + w_R\delta_R,
\end{align}
where, $\delta_L \in[-1,1]$, $\delta_R \in[-1,1]$, $L_T^{N}=L_g + \frac{1}{2}[\bar{L}_{Th}+\underline{L}_{Th}]$, $R_T^{N}=R_g + \frac{1}{2}[\bar{R}_{Th}+\underline{R}_{Th}]$, $w_L=\frac{1}{2}[\bar{L}_{Th}-\underline{L}_{Th}]$ and $w_R=\frac{1}{2}[\bar{R}_{Th}-\underline{R}_{Th}]$. This is the classical real parametric uncertainty modeling, quite commonly found in robust control theory. However, in synthesizing the controller with defined uncertainties in $1/(L_Ts+R_T)$, this representation is quite difficult to handle. Linear fractional transformation (LFT) approach can be utilized to convert the model into an upper LFT, given by $F_U(\mathbf{M},\mathbf{\Delta})$ as follows\cite{robust1} :
\begin{align*}
&\underbrace{\left[
\begin{array}{c|c}
  \mathbf{M_{11}} & \mathbf{M_{12}} \\
  \hline
  \mathbf{M_{21}} & \mathbf{M_{22}}
\end{array}
\right]}_{\mathbf{M}}
=
\dfrac{1}{sL_T^N+R_T^N}
\left[
\begin{array}{c | c}
  \begin{array}{c c}
     -sw_L & -w_R\\
     -sw_L & -w_R
  \end{array} & \begin{array}{c c}
     1\\
     1
  \end{array} \\
  \hline
  \begin{array}{c c}
     -sw_L & -w_R
  \end{array} & 1
 \end{array}
\right],\\
&~s.t.~ \dfrac{1}{sL_T+R_T}=\mathbf{M_{22}} + \mathbf{M_{21}\Delta(I-M_{11}\Delta)^{-1}M_{12}},
\end{align*}
where, $\mathbf{\Delta}=diag(\delta_L, \delta_R)$.
\section{Proposed $\mathcal{H}_{\infty}$-Based Solution}
\subsection{Objectives and Operation of the Control System}
\begin{figure}[t]
\centering
\includegraphics[scale=0.165,trim={0cm 0.0cm 0cm 0cm},clip]{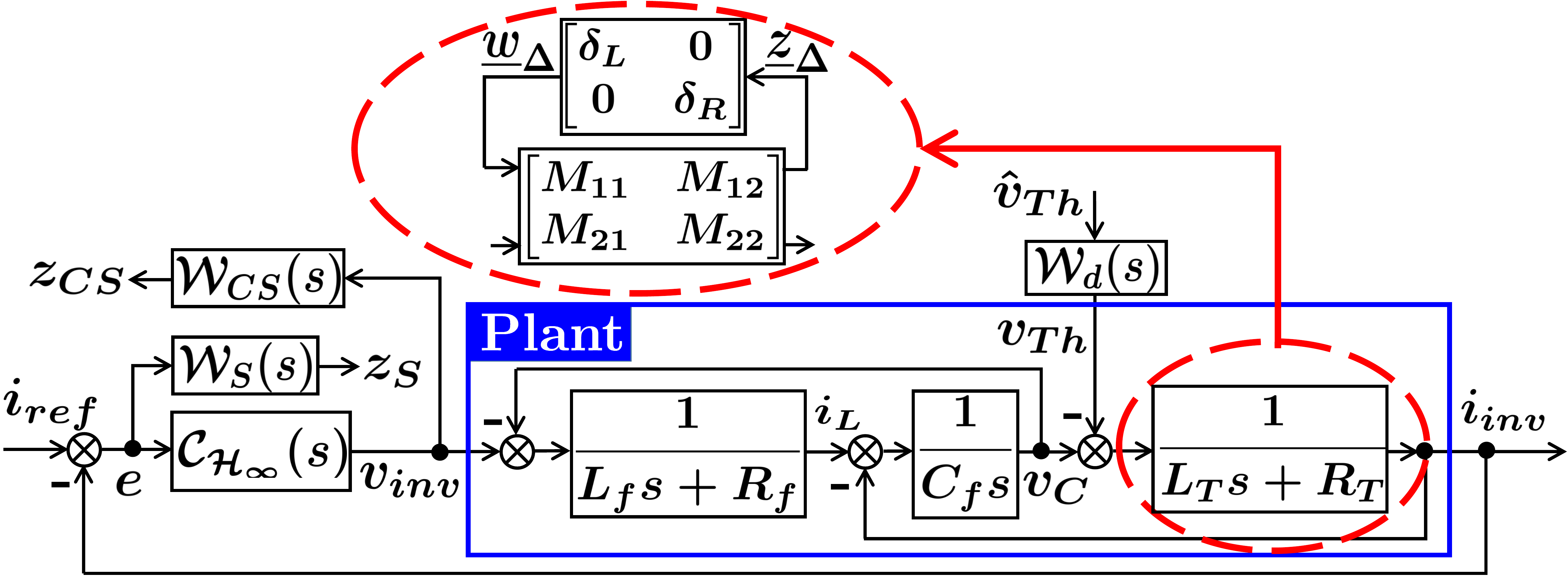}%
\caption{Proposed $\mathcal{H}_{\infty}$-based robust current controller synthesis.}
\label{fig:control1}
\end{figure}

A $P$-$Q$ controlled \textit{gf}VSI operates to inject a pre-specified reference active power, $P_{ref}$, and reactive power, $Q_{ref}$, (defined locally/centrally) into the network by employing a current control strategy \cite{DER_control_Madureira}. An outer `Reference Generation Block' (as shown in Fig.~\ref{fig:circuit}) eventually generates the $i_{ref}$ signal using its $P$-$Q$ set-points and output signals from phase-locked loop (PLL). The expression of $i_{ref}$ is given by
\begin{align}\label{eq:iref}
    i_{ref}(t) = \sqrt{2}\frac{\sqrt{P_{ref}^2+Q_{ref}^2}}{\bar{V}}\sin\bigg({\Tilde{\theta}-\arctan{\frac{Q_{ref}}{P_{ref}}}}\bigg).
\end{align}
A PLL operates with its grid-synchronization technique and generates signals, $\bar{V}$, $\Tilde{\theta}$, containing the RMS value and synchronized phase information of $v_{PCC}$ respectively. In this work, a 1-$\phi$ second order generalized integrator-based synchronous reference frame PLL (SOGI-SRF-PLL) is used \cite{single_phase_pll}. The generated $i_{ref}$ along with the measured $i_{inv}$ are the inputs to the `Control Logic' block as shown in Fig.~\ref{fig:circuit}. The objective here is to design a feedback control law through controller, $\mathcal{C}_{\mathcal{H}_{\infty}}(s)$ as shown in Fig.~\ref{fig:control1}, which generates a control signal, $v_{inv}$, such that, i) $i_{inv}$ tracks $i_{ref}$ with minimum tracking error, ii) effects of $v_{Th}$ on $i_{inv}$ is largely attenuated, iii) $v_{inv}$ satisfies the bandwidth limitations. Moreover, all aforementioned objectives need to be satisfied with model uncertainties. In other words, it is necessary to provide the controller with enough robustness to deal with uncertainties caused by the distribution network. These objectives are derived from acceptable standards on power quality, e.g. IEEE Std 519\cite{ieee2}. 
\subsection{Design Procedure of the $\mathcal{H}_{\infty}$-based Controller}
$\mathcal{H}_{\infty}$-based controller design provides a framework for addressing multiple objectives. Here, the design is based on the system structure illustrated in Fig.~\ref{fig:control1} where user-defined weighting transfer functions, $\mathcal{W}_{{S}}(s)$, $\mathcal{W}_{CS}(s)$, $\mathcal{W}_d(s)$, are selected based on the aforementioned objectives. The guidelines for designing the weighting functions are provided below.
\subsubsection{Selection of $\mathcal{W}_S(s)$}
To shape the sensitivity transfer function, the weighting function, $\mathcal{W}_S(s)$, is introduced so that
\begin{itemize}
    \item the tracking error, $e$, at fundamental frequency is low,
    \item the $LCL$ filter resonance of the VSI is actively damped.
\end{itemize}
$\mathcal{W}_S(s)$ is modeled to have peaks around $\omega_o$ and $LCL$ resonant frequency, $\omega_{r}$, with $2^{nd}$ order roll-off, $k_{S,1}(s)$ and formed as
\begin{align*}
    \mathcal{W}_S(s) = k_{S,1}(s)\dfrac{s^2+2k_{S,2}\zeta\omega_o s+\omega_o^2}{s^2+2\zeta\omega_o s+\omega_o^2} \dfrac{s^2+2k_{S,3}\zeta\omega_{r}s+\omega_{r}^2}{s^2+2\zeta\omega_{r}s+\omega_{r}^2},
\end{align*}
where, $k_{S,2}$ and $k_{S,3}$ are selected to exhibit peaks and $\zeta$ takes care of the off-nominal frequency around the nominal values.
\subsubsection{Selection of $\mathcal{W}_{CS}(s)$}
$\mathcal{W}_{CS}(s)$ is designed to suppress high-frequency control effort to shape the performance of $v_{inv}$. Hence, it is designed as a high-pass filter with cut-off frequency at switching frequency and is ascribed the form:
\begin{align*}
    \mathcal{W}_{CS}(s) = k_{CS}\dfrac{s+k_{CS,1}\omega_o}{s+k_{CS,2}\omega_o}, ~\text{where}~ k_{CS,1}<<k_{CS,2}.
\end{align*}
\subsubsection{Selection of $\mathcal{W}_d(s)$}
$\mathcal{W}_d(s)$ emphasizes the expected disturbances at fundamental and harmonic frequencies imposed by $v_{Th}$ and emphasized by exogenous signal $\hat{v}_{Th}$. It is based on the assumption that the network voltage comprises fundamental and considerable amount of $3^{rd}$, $5^{th}$, $7^{th}$ harmonics \cite{ieee2}. Hence, it is designed by a low-pass filter, $k_d(s)$, with peaks at selected frequencies and is ascribed the form:
\begin{align*}
    \mathcal{W}_d(s) = k_d(s)\prod_{h=1,3,5,7}\dfrac{s^2+2k_{d,h}\zeta h\omega_o s+h^2\omega_o^2}{s^2+2\zeta h\omega_o s+h^2\omega_o^2},
\end{align*}
where, the values of $k_{d,h}$ are selected based on the voltage THD standards recommended in IEEE Std-519 \cite{ieee2}. 
\subsection{Problem Formulation and Resulting Optimal Controller}
\begin{figure}[t]
	\centering
	\subfloat[]{\includegraphics[scale=0.2,trim={0.0cm 0.0cm 0.0cm 0.0cm},clip]{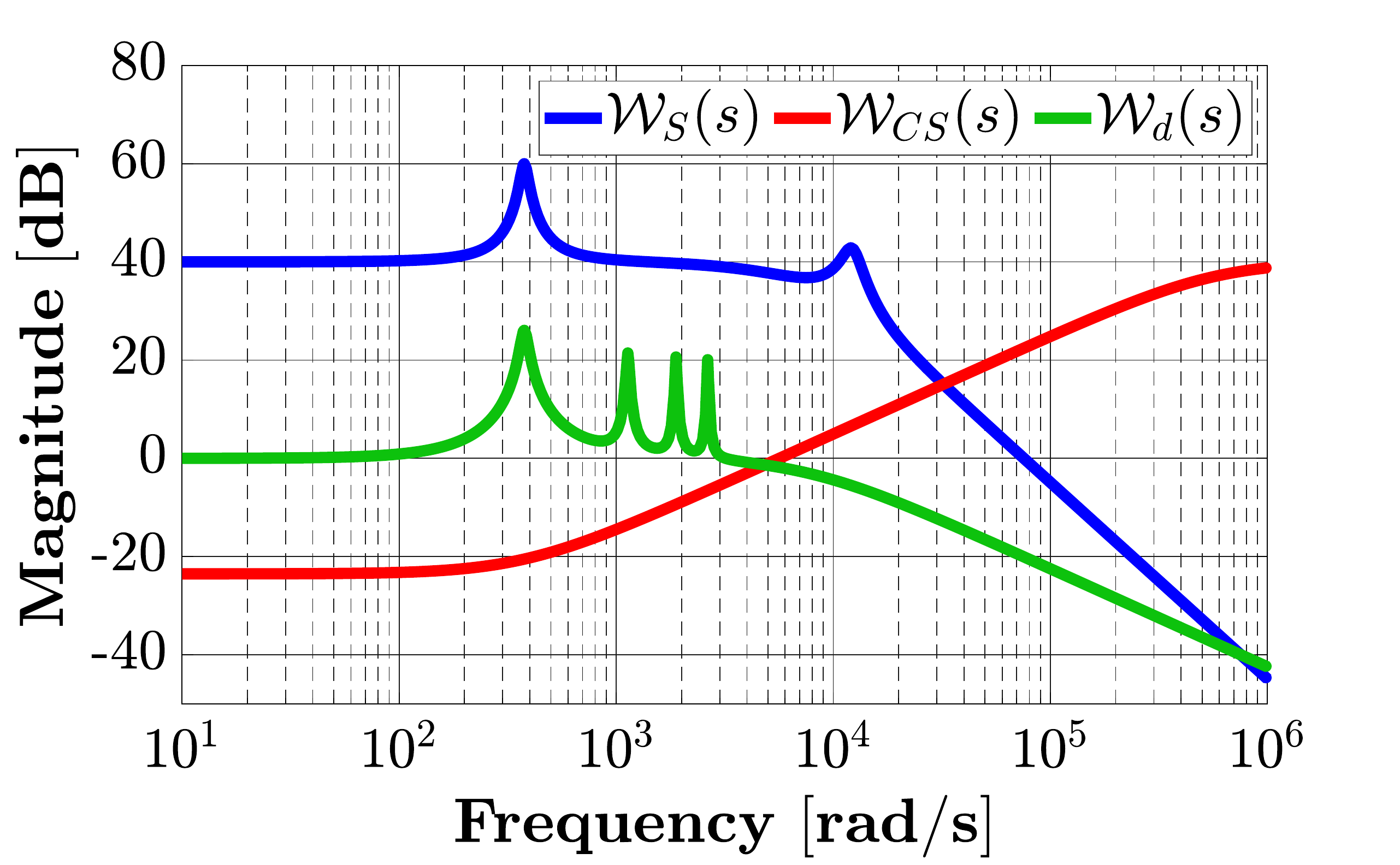}%
	\label{fig:weight}}
	\subfloat[]{\includegraphics[scale=0.17,trim={0.0cm 0.0cm 32.0cm 5cm},clip]{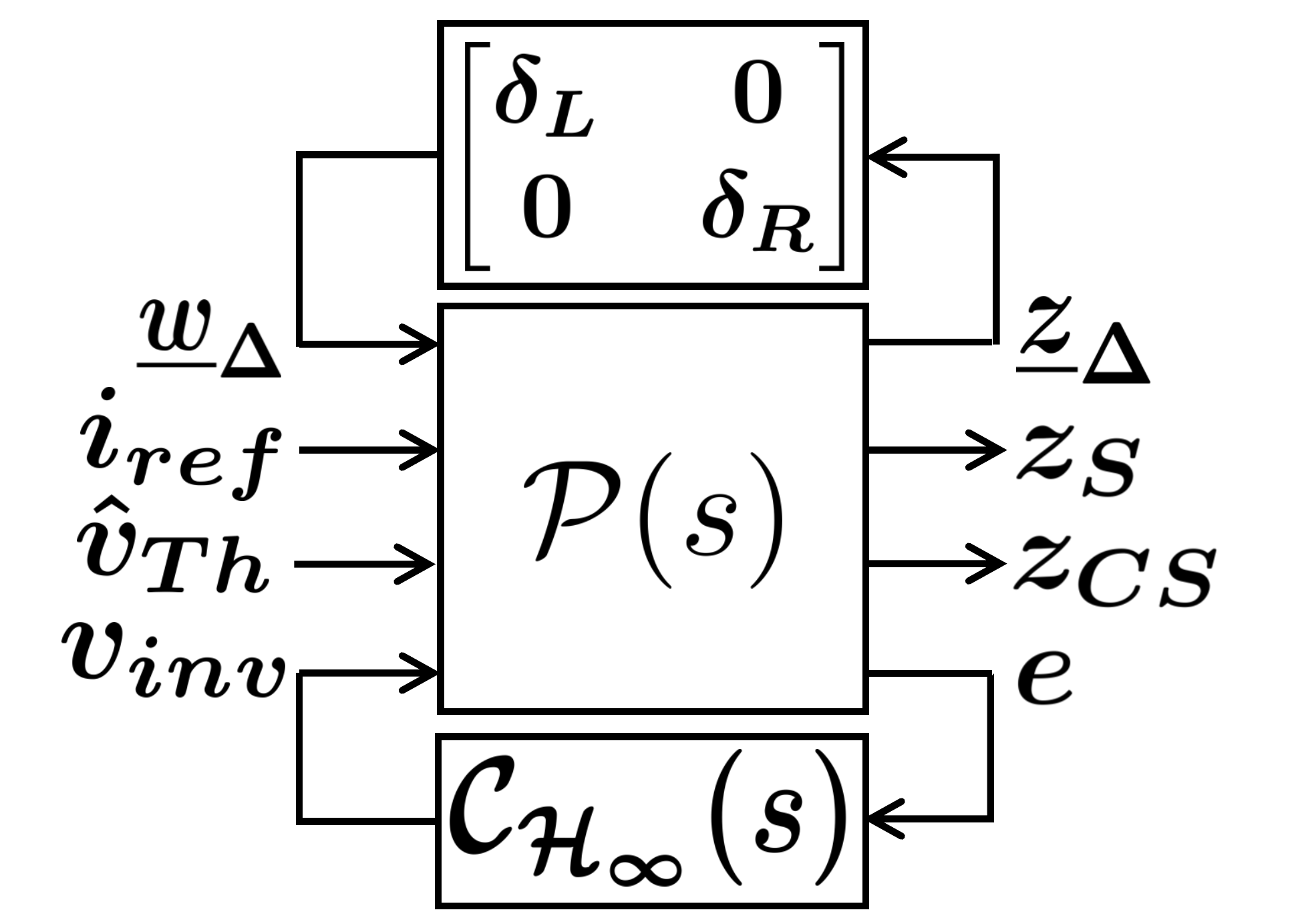}%
	\label{fig:control2}}
	\caption{(a) Bode plots of magnitudes of selected weighting transfer functions, (b) generalized control configuration of Fig.~\ref{fig:control1}.}
	\label{fig:synthesis}
\end{figure}

\begin{figure}[t]
	\centering
	\subfloat[]{\includegraphics[scale=0.24,trim={0.0cm 0.0cm 1.0cm 0cm},clip]{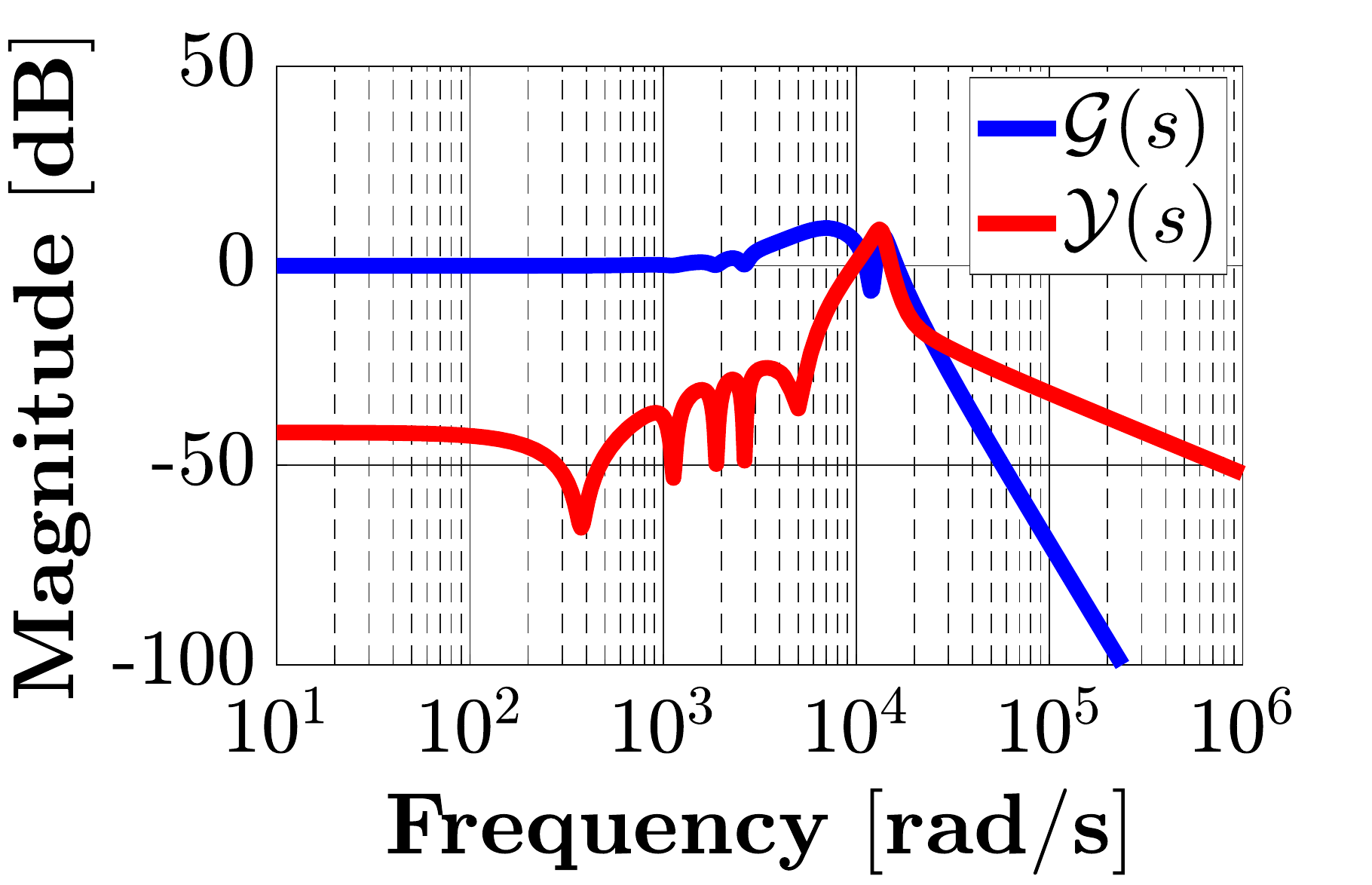}%
		\label{fig:GYmag}}~
    \subfloat[]{\includegraphics[scale=0.24,trim={0.0cm 0.0cm 1.0cm 0.0cm},clip]{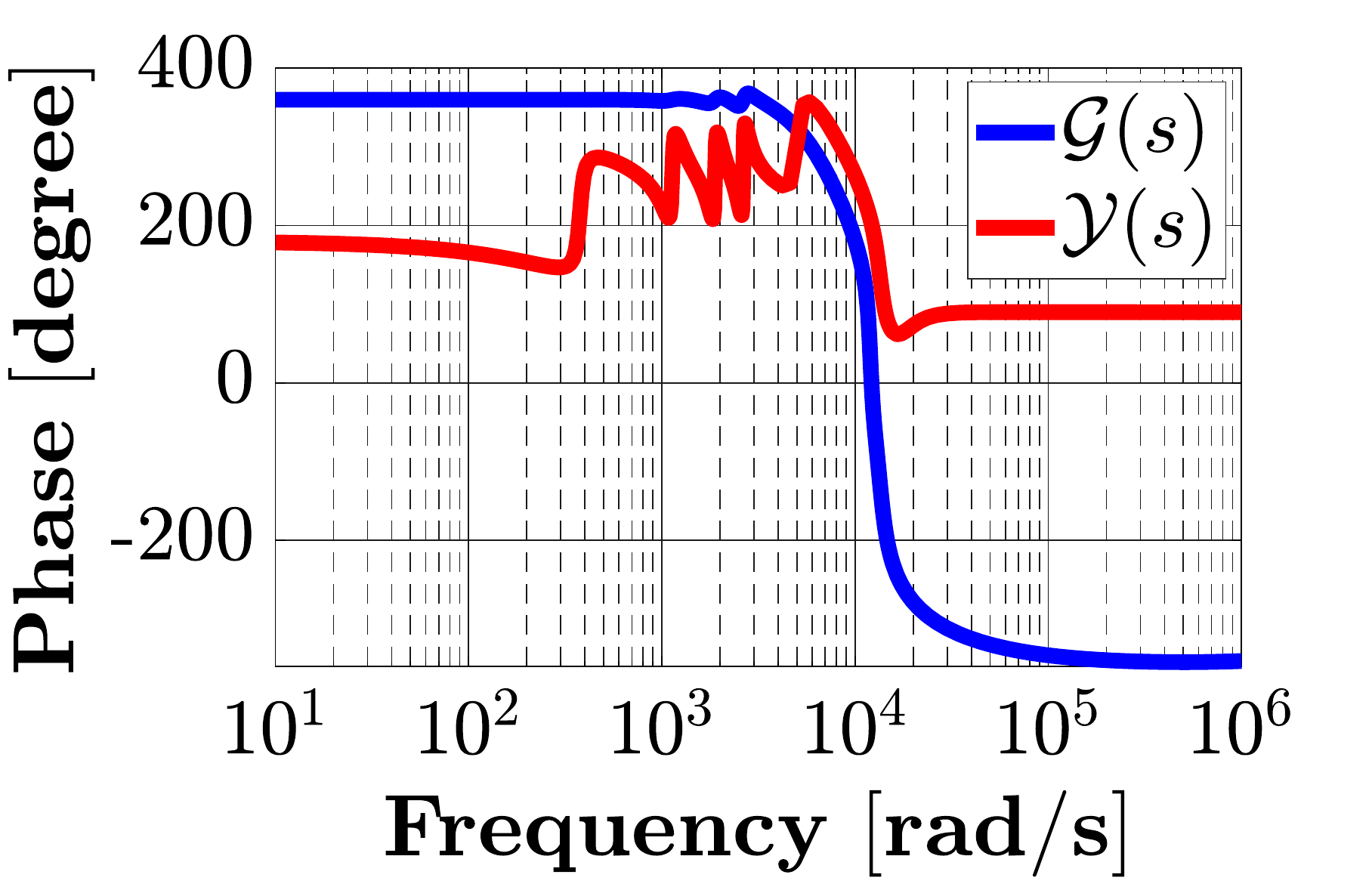}%
		\label{fig:GYphase}}
	\caption{Bode plots, (a) magnitudes of $\mathcal{G}(s)$, $\mathcal{Y}(s)$, (b) phase of $\mathcal{G}(s)$, $\mathcal{Y}(s)$.}
	\label{fig:bode}
\end{figure}

The Bode plots of the selected weighting transfer functions in this work are shown in Fig.~\ref{fig:synthesis}\subref{fig:weight}. The $\mathcal{H}_{\infty}$-based optimal problem is formulated and solved to generate a feedback control law with resulting controller, $\mathcal{C}_{\mathcal{H}_{\infty}}(s)$, stated as:
\begin{align}\label{eq10}
    V_{inv}(s) = \mathcal{C}_{\mathcal{H}_{\infty}}(s)[I_{ref}(s)-I_{inv}(s)].
\end{align}
As a result, the closed loop system equation for the VSI can be found by combining \eqref{eq4} and \eqref{eq10} and can be written as
\begin{align}\label{eq11}
    I_{inv}(s) = \mathcal{G}(s)I_{ref}(s) - \mathcal{Y}(s)V_{Th}(s),
\end{align}
where, 
\begin{align*}
\mathcal{G}(s)=\dfrac{\mathcal{G}_{inv}(s)\mathcal{C}_{\mathcal{H}_{\infty}}(s)}{1+\mathcal{G}_{inv}(s)\mathcal{C}_{\mathcal{H}_{\infty}}(s)},~ \mathcal{Y}(s)=\dfrac{\mathcal{G}_{Th}(s)}{1+\mathcal{G}_{inv}(s)\mathcal{C}_{\mathcal{H}_{\infty}}(s)}.
\end{align*}
\begin{remark}
It is equivalent to state that the optimal controller is required to be designed satisfying the following conditions:
$\mathcal{G}(s)|_{s=j\omega_o} \approx 1\angle 0^o$ and $\mathcal{Y}(s)|_{s=jh\omega_o} << 1$ for $h=1, 3, 5, 7$. 
\end{remark}
The control system of Fig.~\ref{fig:control1} can be realized as a generalized control configuration as shown in Fig.~\ref{fig:synthesis}\subref{fig:control2}\cite{robust1}. It has a generalized MIMO plant, $\mathcal{P}(s)$, containing all nominal models, $\mathbf{M}$, $\mathcal{W}_S(s)$, $\mathcal{W}_{CS}(s)$ and $\mathcal{W}_d(s)$ with exogenous input signal ${w}:=\begin{bmatrix}\underline{w}_{\Delta} & i_{ref} & \hat{v}_{Th} & {v}_{inv}\end{bmatrix}^\top$ and output signals ${z}:=\begin{bmatrix}\underline{z}_{\Delta} & z_S & z_{CS} & e\end{bmatrix}^\top$. The controller, $\mathcal{C}_{\mathcal{H}_{\infty}}(s)$ has input feedback signal, $e$, and output control signal, ${v}_{inv}$. The uncertainty function, $\mathbf{\Delta}$, with input $\underline{z}_{\Delta}$ and output $\underline{w}_{\Delta}$ is represented using upper LFT \cite{robust1}. The goal is to synthesize the stabilizing controller that satisfies the following:
\begin{align}\label{eq12}
    ||T_{w\rightarrow z}||_{\infty} < 1.
\end{align}
By means of \textit{hinfsyn} command of MATLAB robust control toolbox, the synthesis of optimal controller is achieved. Usually $\mathcal{H}_{\infty}$-control algorithms produce controllers of higher order and model reduction becomes essential to design a low order implementable controller. It is achieved by using balanced truncation method removing modes faster than the switching frequency. The resulting controller, $\mathcal{C}_{\mathcal{H}_{\infty}}(s)$, is of the order of 11 which is higher than that of conventional PR controller with harmonic compensators only by 3. Fig.~\ref{fig:bode}\subref{fig:GYmag} and \ref{fig:bode}\subref{fig:GYphase} corroborate the accomplishment of the objectives by the resulting optimal controller with $||T_{w\rightarrow z}||_{\infty}=0.63$. Robust stability is verified by \textit{robstab} command of MATLAB.
\section{Validation and Results}
The computational footprint of the proposed $\mathcal{H}_{\infty}$-based controller, an essential check for validating the performance while implemented in a real low-cost micro-controller board, is discussed here with a brief description of the test system.
\subsection{Controller Hardware-in-the-loop Setup Description}
\begin{figure*}[t]
	\centering
    \includegraphics[scale=0.26,trim={0.0cm 0.0cm 0cm 0.0cm},clip]{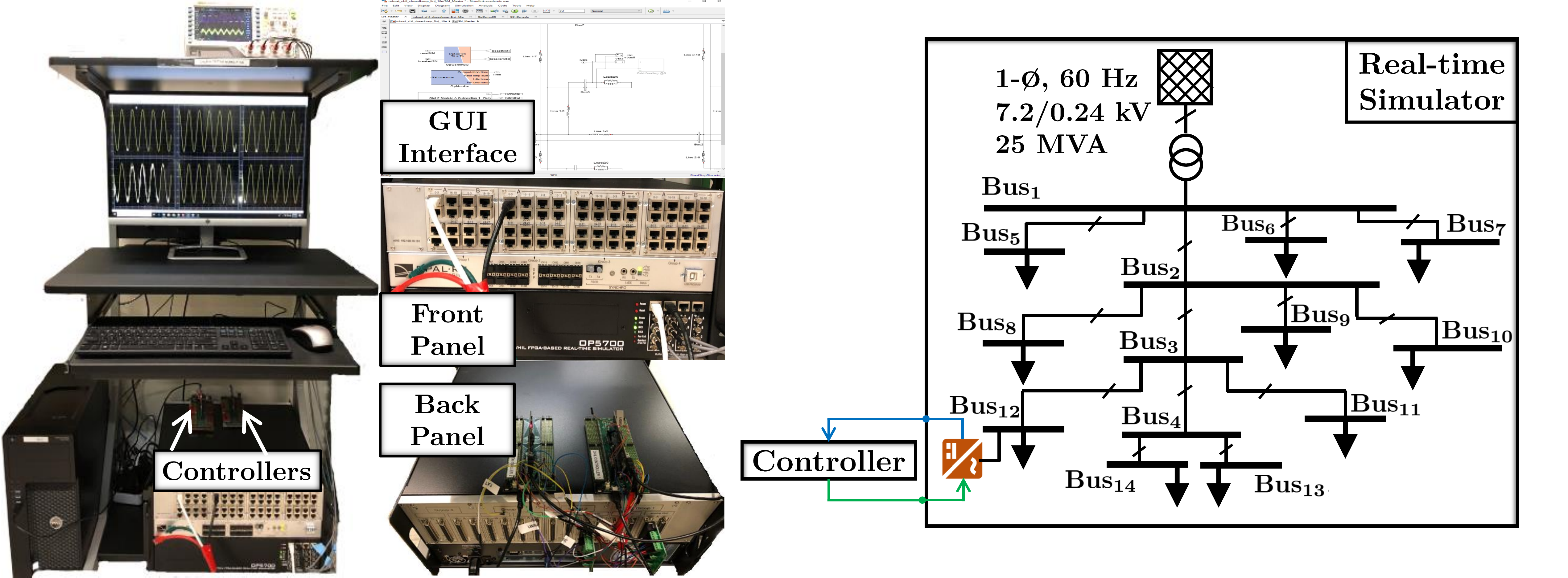}%
	\caption{OPAL-RT based hardware-in-the-loop simulation platform with CIGRE test system and Texas Instruments Delfino TMS320F28379D controller board.}
	\label{fig:chilplatform}
\end{figure*}

Controller hardware-in-the-loop (CHIL)-based simulation studies are conducted on OPAL-RT real-time simulator. North American low voltage distribution feeder from CIGRE Task Force C6.04.02, affiliated with CIGRE Study Committee C6 is emulated inside the OPAL-RT along with the power circuit of a \textit{gf}VSI, connected at $\text{Bus}_{\text{12}}$ with parameters tabulated in Table~\ref{table:data}. Ratings of distribution transformer, loads at each bus and line parameters are provided in \cite{testsystem}. The test system is modified by including sufficient amount of non-linear loads at various buses while respecting the recommended limits of THD mentioned in \cite{ieee2}. The proposed resulting $\mathcal{H}_{\infty}$-based controller is realized on a low-cost Texas-Instruments TMS28379D Delfino controller board as shown in Fig.~\ref{fig:chilplatform}. 
\renewcommand{\arraystretch}{1.2}
\begin{table}[t]
\centering
\caption{GRID AND GRID-FEEDING VSI PARAMETERS UNDER STUDY}
\label{table:data}
\begin{tabular}{|c|c|}
\hline 
\textbf{VSI Parameter} & \textbf{Value}    \\ \hline \hline
Ratings & $240$~V (RMS), $60$~Hz, $11$~kVA, $0.95$~pf \\ \hline
VSI Parameters & $V_{dc}$ = $500$~V, $f_{Sw}$ = $20$~kHz \\ \hline
$LCL$ Filter Parameters & $L_f$ = $2$~mH, $L_g$ = $400$~$\mu$H, $C_f$ = $20~\mu$F \\ \hline \hline \hline
\textbf{Grid Parameter} & \textbf{Value}    \\ \hline \hline
Grid Impedance & $L_{Th}\in[0,0.53]$~mH,~$R_{Th}\in[0,0.05]$~$\Omega$ \\ \hline
\end{tabular}
\end{table}
\subsection{CHIL-based Experimental Result}
\begin{figure}[t]
	\centering
	\subfloat[]{\includegraphics[scale=0.26,trim={3.5cm 6.5cm 0cm 0cm},clip]{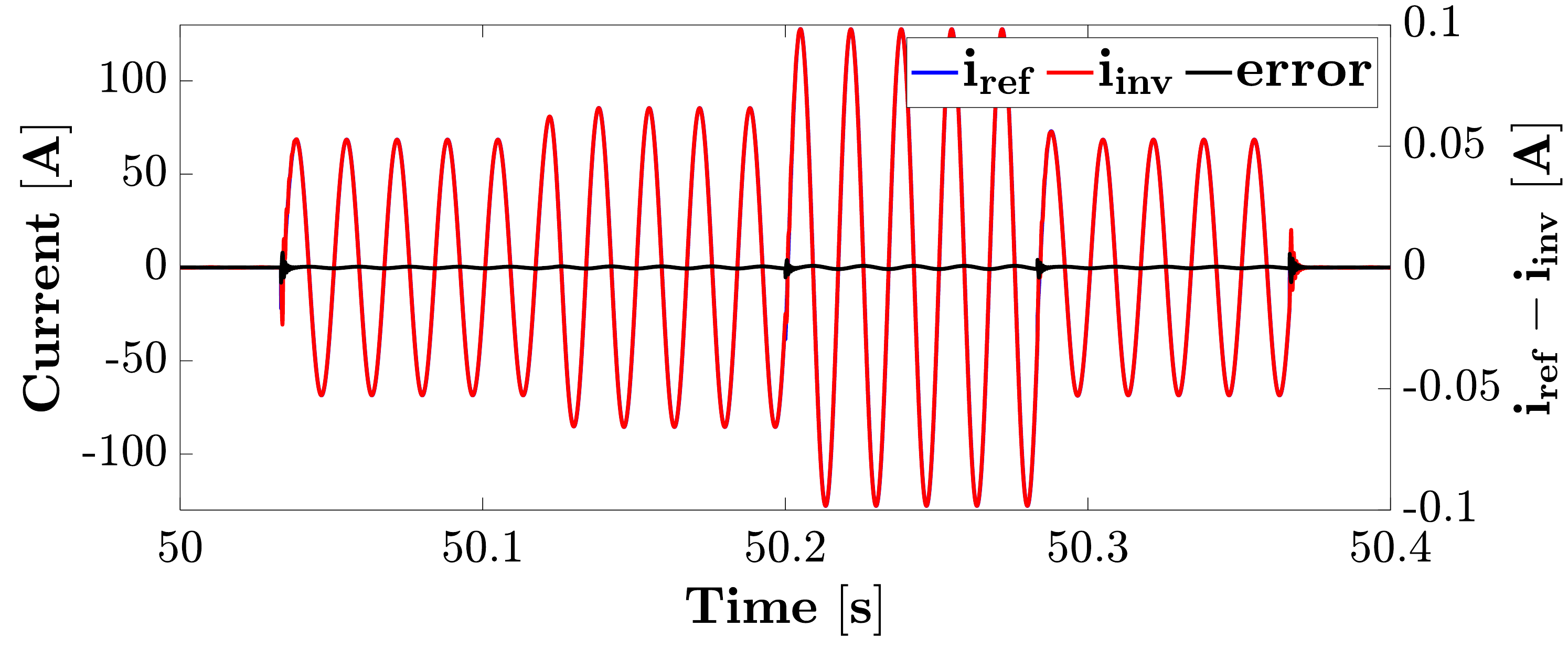}%
		\label{fig:result1}}
		
    \subfloat[]{\includegraphics[scale=0.25,trim={3.5cm 6.5cm -2cm 0cm},clip]{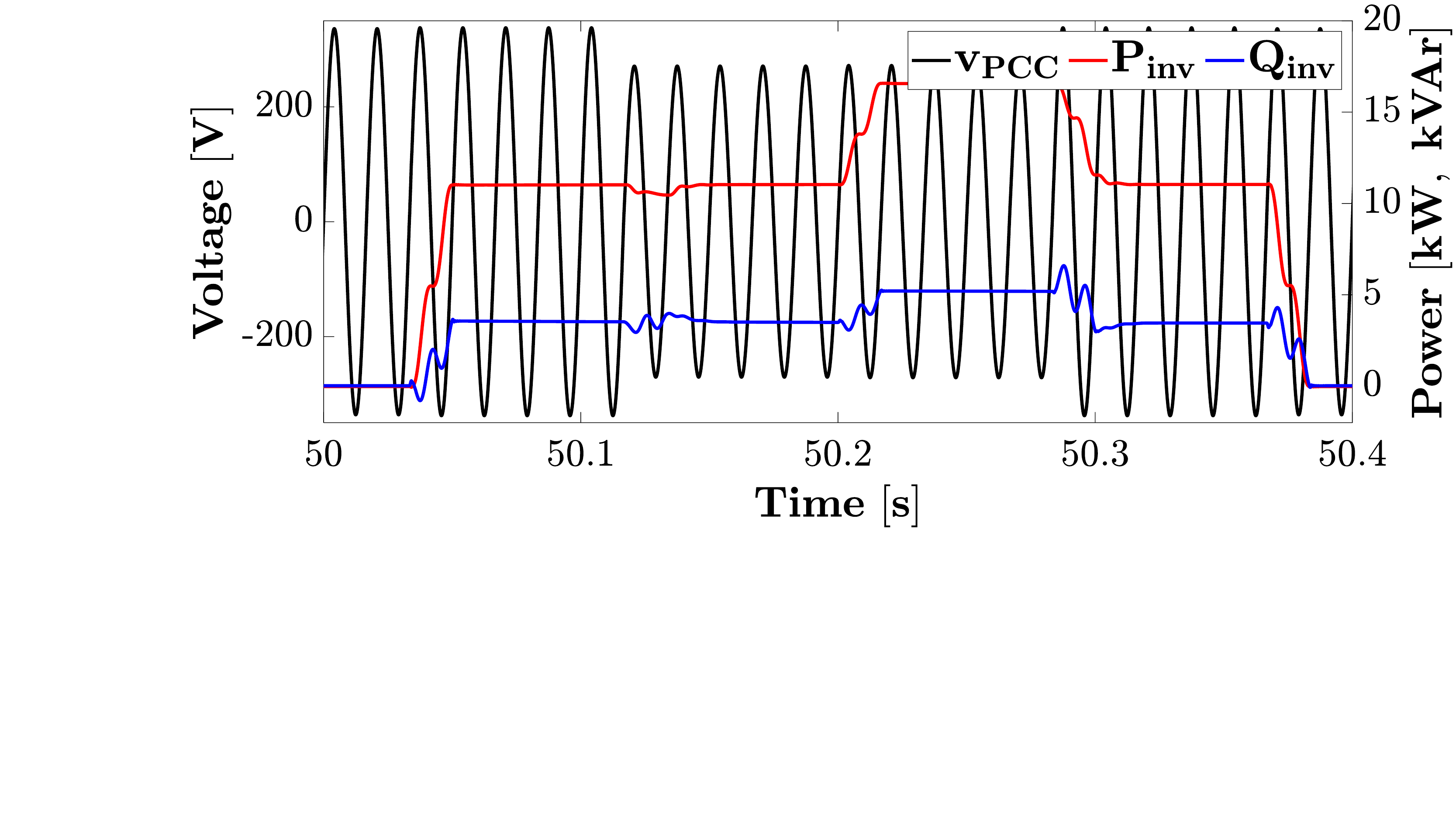}%
		\label{fig:result2}}
	\caption{CHIL simulation results for $\mathtt{CASE}$-$\mathtt{1}$, (a) output current and reference waveform with tracking error, (b) power output and voltage waveform at PCC.}
	\label{fig:result12}
\end{figure}

\begin{figure}[t]
	\centering
	\subfloat[]{\includegraphics[scale=0.26,trim={3.5cm 6.5cm 0cm 0cm},clip]{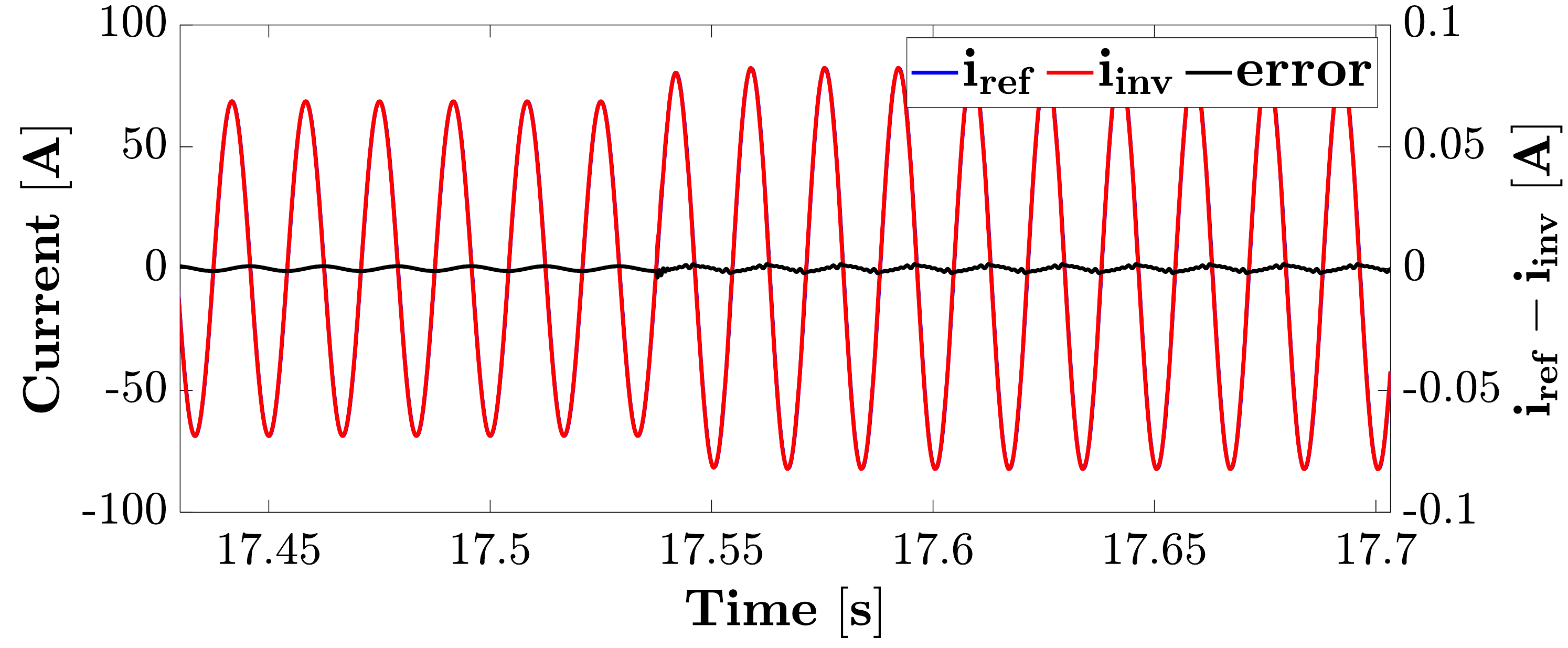}%
		\label{fig:result3}}
		
    \subfloat[]{\includegraphics[scale=0.25,trim={3.5cm 6.5cm -2cm 0cm},clip]{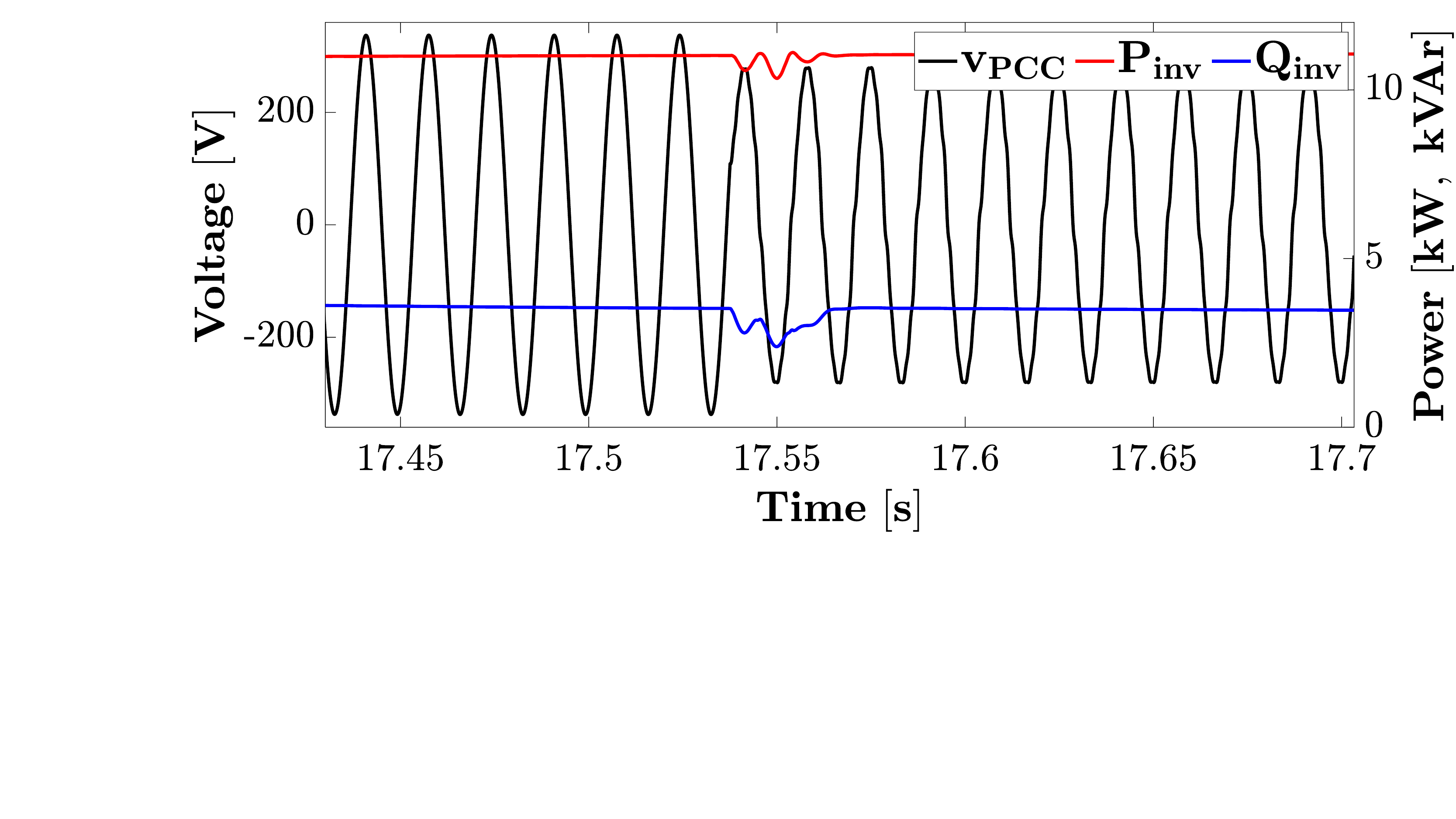}%
		\label{fig:result4}}
	\caption{CHIL simulation results for $\mathtt{CASE}$-$\mathtt{2}$, (a) output current and reference waveform with tracking error, (b) power output and voltage waveform at PCC.}
	\label{fig:result34}
\end{figure}

Two test cases are examined by emulating a sequence of events in the OPAL-RT platform. Test cases are enlisted as:
\begin{itemize}
    \item $\mathtt{CASE}$-$\mathtt{1}$: The VSI is initially in no-load condition. At $t=50.03~s$, there is a transition from no-load to full-load and stays until $t=50.2~s$. $20$\% under-voltage (of nominal) occurs at $t=50.12~s$ and stays until $t=50.28~s$ when voltage revives to nominal value. During this interval, VSI is overloaded by $150$\% from $t=50.2~s$. At $t=50.28~s$, the VSI is switched to rated load condition until $t=50.36~s$ when it switches over to no load condition.
    \item $\mathtt{CASE}$-$\mathtt{2}$: The VSI is operating in rated loading with a sudden jump of equivalent grid inductance from $0$~mH to $0.53$~mH at $t=17.53~s$ while maintaining same loading.
\end{itemize}
The results for $\mathtt{CASE}$-$\mathtt{1}$ are shown in Fig.~\ref{fig:result12}. It is clearly observed that the output current waveforms of VSI, $i_{inv}$, is following $i_{ref}$ with minimal tracking error as shown in Fig.~\ref{fig:result12}\subref{fig:result1}. The current and power outputs are maintained during the sudden changes of $v_{PCC}$ as illustrated in Fig.~\ref{fig:result12}\subref{fig:result2}. Thus, the performance of the proposed controller of reference tracking and disturbance rejection are validated in this case studies. The results for $\mathtt{CASE}$-$\mathtt{2}$ are shown in Fig.~\ref{fig:result34}\subref{fig:result3} and Fig.~\ref{fig:result34}\subref{fig:result4}. It is observed that the performance of the proposed current controller is sufficiently robust to a substantial amount of variations in equivalent grid impedance and that validates the robust performance of the proposed controller.
\subsection{Performance Comparison}
For the purpose of performance comparison, PR-based current controller with PCC voltage feed-forward is considered for the CHIL simulation. Reference \cite{yazdani} provides an elaborated guidelines for designing the current controller with sufficient gain and phase margins in order to possess a fair comparative study. The current loop with PR controller is designed to have PM $\geq 45^{\circ}$ and GM $ \geq 40$~dB with bandwidth of $1.5$~kHz. Similar to $\mathtt{CASE}$-$\mathtt{1}$ and $\mathtt{CASE}$-$\mathtt{2}$, $\mathtt{CASE}$-$\mathtt{3}$ study is conducted for comparison study of proposed $\mathcal{H}_{\infty}$-based controller with PR controller in this work. The case study is as follows:
\begin{itemize}
    \item $\mathtt{CASE}$-$\mathtt{3}$: The VSI is operating at rated loading with minimal grid impedance (stiff grid) until $t=17.53~s$ when there is a sudden jump of grid inductance from $0.1$~mH to $0.53$~mH (weak grid). Moreover, at $t=17.62~s$ VSI jumps to $150\%$ loading condition with weak grid. 
\end{itemize}
\begin{figure}[t]
	\centering
	\subfloat[]{\includegraphics[scale=0.26,trim={3.5cm 6.5cm 0cm 0cm},clip]{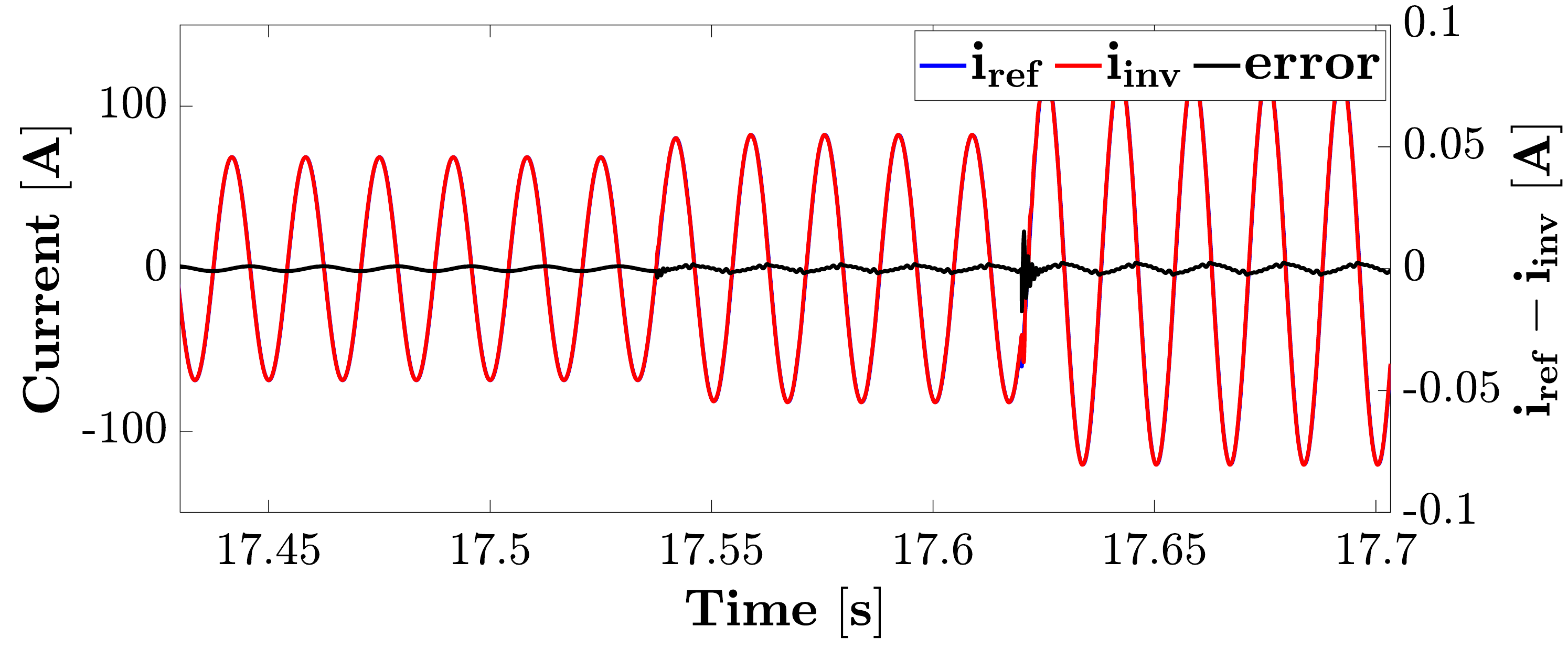}%
		\label{fig:result5}}
		
    \subfloat[]{\includegraphics[scale=0.26,trim={3.5cm 6.5cm 0cm 0cm},clip]{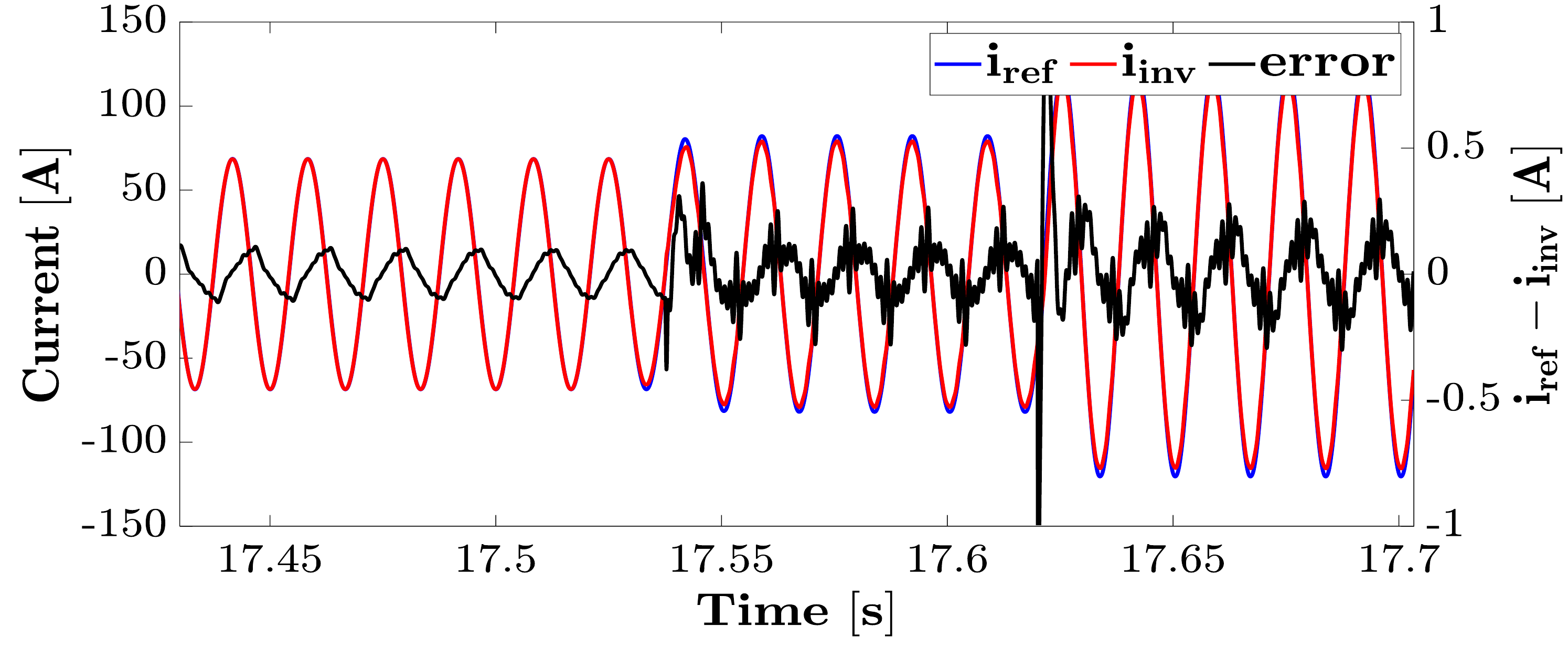}%
		\label{fig:result7}}
	\caption{CHIL simulation results for $\mathtt{CASE}$-$\mathtt{3}$ with output current and reference waveform, tracking error, (a) $\mathcal{H}_{\infty}$-based controller, (b) PR controller.}
	\label{fig:result57}
\end{figure}
The results for $\mathtt{CASE}$-$\mathtt{3}$ are shown in Fig.~\ref{fig:result57}. It is observed that the output current of VSI, $i_{inv}$, is following $i_{ref}$ with large tracking error in PR controller once the grid stiffness weakens as shown in Fig.~\ref{fig:result12}\subref{fig:result5} and Fig.~\ref{fig:result12}\subref{fig:result7} respectively. The results substantiate the fact that the proposed $\mathcal{H}_{\infty}$-based controller exhibits superior robustness in performance than the PR controller at the cost of increasing the order of controller only by $3$ and no additional sensor requirements. 
\section{Conclusions}
This article demonstrates the design and implementation of a robust current controller for single-phase \textit{gf}VSI. The uncertainty in grid impedance is modeled explicitly to leverage the robustness in performance of the controller. $\mathcal{H}_{\infty}$-based controller design is followed and the required objectives for the optimal controller are discussed which leads to the final optimal controller. OPAL-RT based CHIL studies are conducted to verify the viability of the resulting controller. Moreover performance with classical PR based controller are compared to highlights the superiority of the proposed controller. 
\section*{Acknowledgment}
The authors acknowledge Advanced Research Projects Agency-Energy (ARPA-E) for supporting this research through the project titled ``Rapidly Viable Sustained Grid'' via grant no. DE-AR0001016.


\end{document}